\documentclass[aps,prl,twocolumn,preprintnumbers,superscriptaddress,amsmath,amssymb]{revtex4}

\usepackage{graphicx}
\usepackage{subfigure}
\usepackage{epsfig}
\usepackage{xcolor}
\usepackage{dcolumn}
\usepackage{bm}
\usepackage{ulem}
\usepackage{color}
\usepackage{amsmath}
\usepackage{amsfonts}
\usepackage{amssymb}

\usepackage{braket}

\begin{document}

\title{Revealing the two-dimensional electronic structure and  anisotropic superconductivity in a natural van der Waals superlattice  (PbSe)$_{1.14}$NbSe$_2$}

\author{Haoyuan Zhong}
\altaffiliation{These authors contributed equally to this work}
\affiliation
{State Key Laboratory of Low-Dimensional Quantum Physics and Department of Physics, Tsinghua University, Beijing 100084, P. R. China}

\author{Hongyun Zhang}
\altaffiliation{These authors contributed equally to this work}
\affiliation
{State Key Laboratory of Low-Dimensional Quantum Physics and Department of Physics, Tsinghua University, Beijing 100084, P. R. China}

\author{Haoxiong Zhang}
\altaffiliation{These authors contributed equally to this work}
\affiliation
{State Key Laboratory of Low-Dimensional Quantum Physics and Department of Physics, Tsinghua University, Beijing 100084, P. R. China}
 
\author{Ting Bao}
\affiliation
{State Key Laboratory of Low-Dimensional Quantum Physics and Department of Physics, Tsinghua University, Beijing 100084, P. R. China}

\author{Kenan Zhang}
\affiliation
{State Key Laboratory of Low-Dimensional Quantum Physics and Department of Physics, Tsinghua University, Beijing 100084, P. R. China}

\author{Shengnan Xu}
\affiliation
{State Key Laboratory of Low-Dimensional Quantum Physics and Department of Physics, Tsinghua University, Beijing 100084, P. R. China}

\author{Laipeng Luo}
\affiliation
{State Key Laboratory of Low-Dimensional Quantum Physics and Department of Physics, Tsinghua University, Beijing 100084, P. R. China}

\author{Awabaikeli Rousuli}
\affiliation
{State Key Laboratory of Low-Dimensional Quantum Physics and Department of Physics, Tsinghua University, Beijing 100084, P. R. China}

\author{Wei Yao}
\affiliation
{State Key Laboratory of Low-Dimensional Quantum Physics and Department of Physics, Tsinghua University, Beijing 100084, P. R. China}

\author{Jonathan D. Denlinger}
\affiliation
{Advanced Light Source, Lawrence Berkeley National Laboratory, Berkeley, California 94720, USA}

\author{Yaobo Huang}
\affiliation
{Shanghai Synchrotron Radiation Facility, Shanghai Advanced Research Institute, Chinese Academy of Sciences, Shanghai 201204, P. R. China}

\author{Yang Wu}
\affiliation
{Tsinghua-Foxconn Nanotechnology Research Center, Tsinghua University, Beijing 100084, P. R. China}
\affiliation
{College of Math and Physics, Beijing University of Chemical Technology, Beijing 100029, P. R. China}

\author{Yong Xu}
\affiliation
{State Key Laboratory of Low-Dimensional Quantum Physics and Department of Physics, Tsinghua University, Beijing 100084, P. R. China}
\affiliation
{Frontier Science Center for Quantum Information, Beijing 100084, P. R. China}

\author{Wenhui Duan}
\affiliation
{State Key Laboratory of Low-Dimensional Quantum Physics and Department of Physics, Tsinghua University, Beijing 100084, P. R. China}
\affiliation
{Frontier Science Center for Quantum Information, Beijing 100084, P. R. China}

\author{Shuyun Zhou}
\altaffiliation{Correspondence should be sent to syzhou@mail.tsinghua.edu.cn}
\affiliation
{State Key Laboratory of Low-Dimensional Quantum Physics and Department of Physics, Tsinghua University, Beijing 100084, P. R. China}
\affiliation
{Frontier Science Center for Quantum Information, Beijing 100084, P. R. China}

\date{\today}

\begin{abstract}

{\bf Van der Waals superlattices are important for tailoring the electronic structures and properties of layered materials. Here we report the superconducting  properties and electronic structure of a natural van der Waals superlattice (PbSe)$_{1.14}$NbSe$_2$. Anisotropic superconductivity with a transition temperature  $T_c$ = 5.6 $\pm$ 0.1 K, which is higher than monolayer NbSe$_2$, is revealed  by transport measurements on high-quality samples.  Angle-resolved photoemission spectroscopy (ARPES) measurements reveal the two-dimensional electronic structure and a charge transfer of 0.43 electrons per NbSe$_2$ unit cell from the blocking PbSe layer.  In addition, polarization-dependent ARPES measurements reveal a significant circular dichroism with opposite contrast at K and K$^\prime$ valleys, suggesting a significant spin-orbital coupling and distinct orbital angular momentum. Our work suggests natural van der Waals superlattice as an effective pathway for achieving intriguing properties distinct from both the bulk and monolayer samples.}

\end{abstract}

\maketitle
The interlayer coupling plays a critical role in determining the fundamental properties of layered materials \cite{wangqh,KisRev2017}, in particular for transition metal dichalcogenides (TMDCs) which exhibit intriguing layer-dependent properties with potential applications in nanoelectronics, optoelectronics, and valleytronics  \cite{wangqh,xuvalley}. Reducing the dimensionality to the two-dimensional (2D) limit leads to distinctive properties from the bulk materials, such as indirect-to-direct bandgap transition \cite{HeinzMoS2,WangFMoS2NL}, Ising superconductivity with an in-plane upper critical field exceeding the Pauli limit \cite{Xi2015Ising,YeMoS2Ising,wang2019type,DingZhangIsing}. So far, the dimensionality control has been mostly achieved by mechanical exfoliation of the bulk crystal or growing monolayer film.  However, atomically thin films in particular metallic films such as NbSe$_2$ \cite{Xi2015Ising,nbstable_nc}, WTe$_2$ \cite{wtflake1} are often unstable under ambient conditions, and sometimes this can  lead to degraded material properties. 
 	
 Misfit layer compounds (MLCs), natural van der Waals superlattices formed by alternating rock-salt monochalcogenide blocking layer and active TMDC layer along the $c$-axis direction \cite{wiegers1992structures,misfit1996,dolotko2020unprecedented}, provide another pathway to tailor the dimensionality and electronic properties of layered materials.  By inserting the blocking layer, the interlayer coupling between neighboring active TMDC monolayers is reduced or even removed, while at the same time, the active monolayers are protected by the blocking layers.  In this way, 2D properties can be achieved in a bulk van der Waals superlattice with high sample quality and robust sample stability as has been reported in NbS$_2$/Ba$_3$NbS$_{5}$ \cite{nbs2science}.  
So far, most studies on MLCs have focused on the growth, characterization and transport measurements \cite{nbs2science,tase2,bai2018superconductivity,bai2020multi,bai2018monolayer,wang2016structure}, while the effect of the blocking layer on the material properties \cite{leriche2021misfit,PSBS2022PRM}, as well as   the tailored electronic structure of MLC as compared to bulk and monolayer counterparts are only rarely investigated.  

 In this work, we report the tailored electronic structure of a natural van der Waals superlattice (PbSe)$_{1.14}$NbSe$_2$, which shows anisotropic superconductivity with a transition temperature  $T_c$ = 5.6 K, which is higher than previous reports on the same compound \cite{Oosawa1992sc,Auriel1993Structure,psns2k,auriel1995electrical,Grosse2016Superconducting} as well as monolayer samples \cite{Xi2015Ising,IS2,IS3,IS4,nbse2yw}. The 2D electronic structure  is revealed by angle-resolved photoemission spectroscopy (ARPES) measurements, and a charge transfer  from PbSe to NbSe$_2$  by 0.43 electrons per NbSe$_2$  unit cell is extracted from the size of the Fermi pockets.  In addition, polarization-dependent ARPES measurements reveal circular dichroism with opposite contrast for the K and K$^\prime$ valleys, indicating distinct orbital angular momentum (OAM) and selective excitation of K and K$^\prime$ valleys. Our work demonstrates  natural van der Waals superlattice as an effective pathway for tailoring the electronic structure of layered materials, and for achieving novel properties with potential applications.

\begin{figure*}[htbp]
  \includegraphics[]{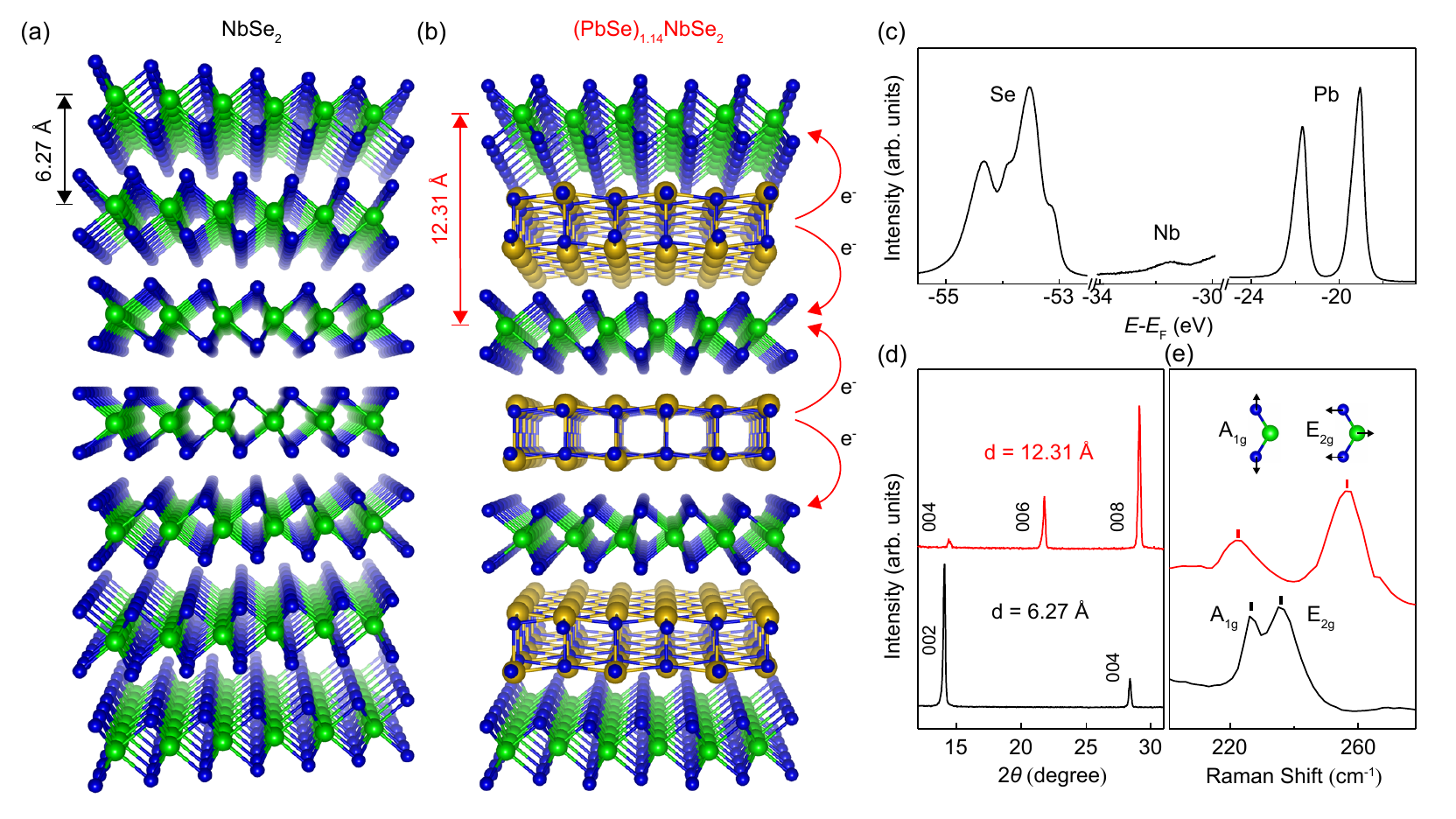}
  \caption{Schematics for van der Waals superlattice  (PbSe)$_{1.14}$NbSe$_2$ and sample characterization. (a,b) Crystal structures of 2H-NbSe$_2$ single crystal and (PbSe)$_{1.14}$NbSe$_2$. The red arrows indicate the charge transfer from PbSe to NbSe$_2$ layers. (c) XPS spectrum shows characteristic Nb, Se and Pb peaks measured under $h\nu$ = 110 eV. (d)  XRD measurements on (PbSe)$_{1.14}$NbSe$_2$ (red) and the NbSe$_2$ single crystal (black). (e) Raman spectra of (PbSe)$_{1.14}$NbSe$_2$ (red) and NbSe$_2$ (black).}
  \label{F1}
\end{figure*}

Figure 1(a,b) shows the crystal structures of bulk NbSe$_2$ and (PbSe)$_{1.14}$NbSe$_2$.  High-quality (PbSe)$_{1.14}$NbSe$_2$ single crystals were grown by chemical vapor transport (CVT) methods \cite{Oosawa1992sc,Auriel1993Structure}, where the chemical stoichiometry is determined by the ratio of the mismatched $a$-axis periodicities \cite{misfit1996}. The as-grown samples were characterized by X-ray diffraction (XRD, sensitive to the out-of-plane crystallinity), Laue diffraction (sensitive to in-plane crystal structure) and Raman spectra (sensitive to the lattice vibrations), and a systematic optimization of the growth conditions including precursors, transport agent and temperature gradient is performed to obtain high-quality samples with good reproducibility (see Fig.~S1, S2 and Table S1 for more details of sample optimization in Supplemental Material \cite{supp}). The optimization of the growth condition can also be applied to other MLCs (see Fig.~S3 for more details of growth of other MLCs in Supplemental Material \cite{supp}). The high-quality samples pave an important step for transport measurements which reveal a higher superconducting transition temperature, and for electronic structure investigations presented below.
 
 The successful insertion of PbSe blocking layers into NbSe$_2$ is confirmed by characteristic Nb, Se, and Pb peaks in the X-ray photoemission spectroscopy (XPS) spectrum measured on the cleaved surface (Fig.~1(c)).  XRD measurements show that the insertion of PbSe blocking layers results in an increase of the interlayer spacing from 6.27 $\pm$ 0.01 \AA~ in the bulk NbSe$_2$ (black curve in Fig.~1(d)) to 12.31 $\pm$ 0.01 \AA~ in the (PbSe)$_{1.14}$NbSe$_2$ (red curve in Fig.~1(d)). The increasing interlayer spacing leads to a reduced interlayer coupling and a change of the vibrational modes. A comparison of the Raman spectra in Fig.~1(e) shows that from NbSe$_2$ to (PbSe)$_{1.14}$NbSe$_2$, the A$_{1g}$ peak shifts from 226.4 cm$^{-1}$ to 222.4 cm$^{-1}$ while the E$_{2g}$ peak from 236.4 cm$^{-1}$ to  256.2 cm$^{-1}$. The A$_{1g}$ and E$_{2g}$ modes involve Se atoms moving perpendicular and parallel to the Nb plane, and therefore they are particularly sensitive to the interlayer coupling \cite{Xiaoxiang2015Strongly}. The redshift of the A$_{1g}$ peak and blueshift of E$_{2g}$ is similar to what is observed from bulk to monolayer MoS$_2$ \cite{raman_ms1,raman_ms2}, indicating weakening of the interlayer coupling in (PbSe)$_{1.14}$NbSe$_2$  (note that the blueshift of E$_{2g}$ may also attribute 
to electron doping from PbSe \cite{2022zsynbse2,das2008monitoring}). 

\begin{figure*}[htbp]
  \includegraphics[]{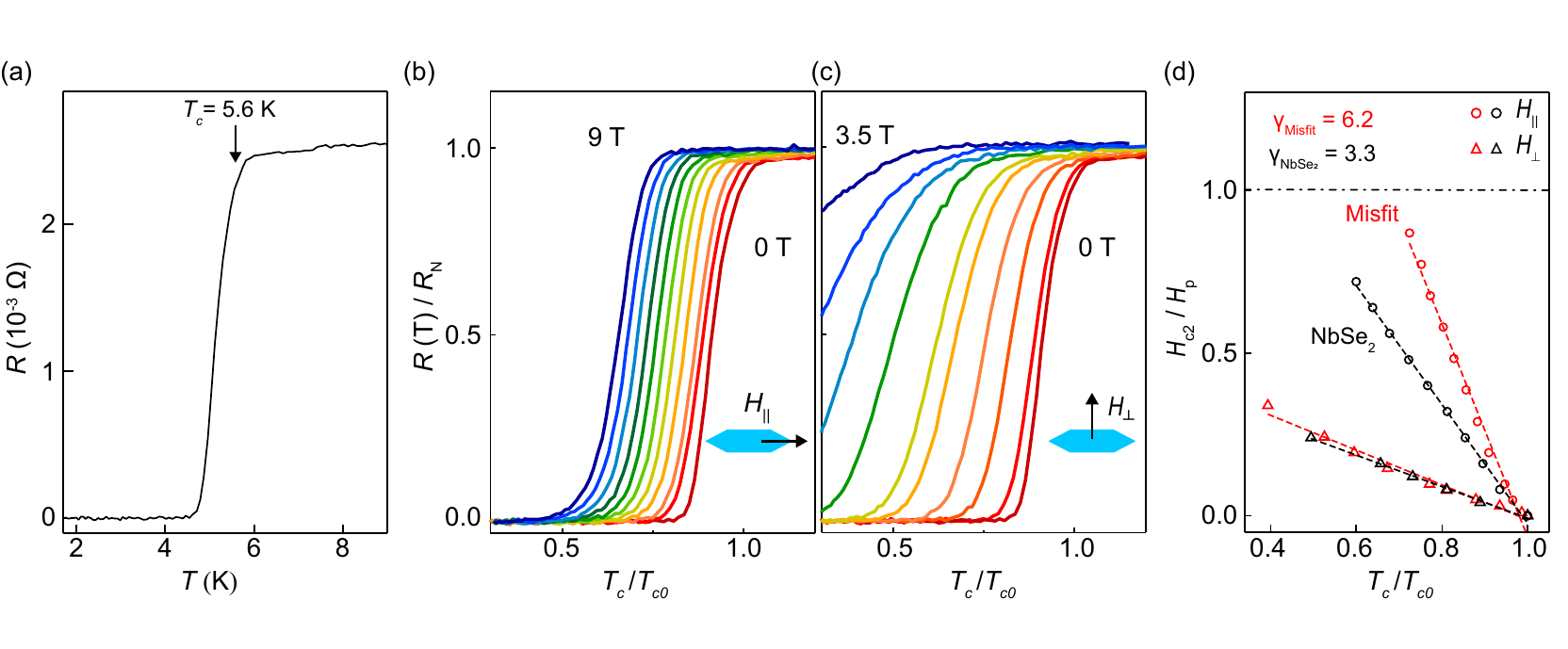}
  \caption{Anisotropic superconductivity with an enhanced in-plane upper critical field. (a) Temperature dependent resistance of (PbSe)$_{1.14}$NbSe$_2$ with transition temperature $T_c$ of 5.6 K, which is defined as the 90$\%$ resistance of the normal state. (b) Normalized resistance upon application of in-plane magnetic fields for (PbSe)$_{1.14}$NbSe$_2$. The in-plane magnetic fields are 0.0, 0.5, 1.0, 2.0, 3.0, 4.0, 5.0, 6.0, 7.0, 8.0 and 9.0 T from red to blue curves and $T_{c0}$ is transition temperature without magnetic field.  (c) Normalized resistance upon application of out-of-plane magnetic fields for (PbSe)$_{1.14}$NbSe$_2$. The out-of-plane magnetic fields are 0.0, 0.1, 0.3, 0.5, 0.8, 1.0, 1.5, 2.0, 2.5 and 3.5 T from red to blue curves. (d) Extracted upper critical magnetic fields $H_{c2}$ for $H_\parallel$ (circles) and $H_\perp$ (triangles) as a function of transition temperature.  Red markers correspond to (PbSe)$_{1.14}$NbSe$_2$, and black markers correspond to NbSe$_2$. Red and black dashed lines are Lawrence-Doniach model  fit.}
  \label{F2}
\end{figure*}

\label{Tex2}

 The as-grown (PbSe)$_{1.14}$NbSe$_2$ exhibits anisotropic superconductivity, which is revealed by transport measurements (Fig.~2).  The optimized sample shows a higher transition temperature $T_c$ = 5.6 $\pm$ 0.1 K (shown in Fig.~2(a)) compared to previously reported $T_c$ values ranging from 1.1 to 2.5 K  \cite{Oosawa1992sc,Auriel1993Structure,psns2k,auriel1995electrical,Grosse2016Superconducting}, together with a sharper transition, which possibly originates from the improved sample quality (see Fig.~S4 for more details of sample with different quality in Supplemental Material \cite{supp}). 
	In addition, the superconductivity shows an anisotropic response to applied magnetic fields - the transition temperature is much less sensitive to the application of in-plane magnetic fields ($H_{\parallel}$, Fig.~2(b)) as compared to out-of-plane magnetic fields ($H_\perp$, Fig.~2(c)). Figure 2(d) shows the extracted upper critical magnetic fields $H_{c2}$ (normalized by Pauli limit $H_P$), and a comparison of (PbSe)$_{1.14}$NbSe$_2$ (red markers) with bulk NbSe$_2$ (black markers, see Fig.~S5 for  raw data in Supplemental Material \cite{supp}) shows that the superconducting anisotropy is enhanced. To quantify the enhanced anisotropy of superconductivity, the anisotropy factor is defined by $\gamma = H_{c2,\parallel}(0)/H_{c2,\perp}(0)$, where $H_{c2,\parallel}(0)$ and $H_{c2,\perp}(0)$ are extracted by Lawrence-Doniach model fit \cite{Trahms2018Superconductive} (shown as dashed lines in Fig.~2(d)). The anisotropy factor is enhanced from $3.3$ for bulk NbSe$_2$ to $6.2$ for (PbSe)$_{1.14}$NbSe$_2$, which is smaller than the Ising superconductivity in monolayer NbSe$_2$ \cite{Xi2015Ising} with a $T_c$ of 0.9 - 3.7 K \cite{Xi2015Ising,IS2,IS3,IS4,nbse2yw}. The enhanced anisotropic is consistent with the increasing interlayer spacing suggested by XRD and Raman measurements in Fig.~1(d,e) which enhances two-dimensional features. We note that similar anisotropy enhancement is also observed in (LaSe)$_{1.14}$NbSe$_2$ \cite{lase2} and [(SnSe)$_{1+\delta}$]$_m$ (NbSe$_2$)$_1$ \cite{Trahms2018Superconductive}, which also confirms that the blocking layer effectively decouples the NbSe$_2$ layer and results in the enhancement of 2D electronic properties.

	\begin{figure*}[htbp]
  \includegraphics[]{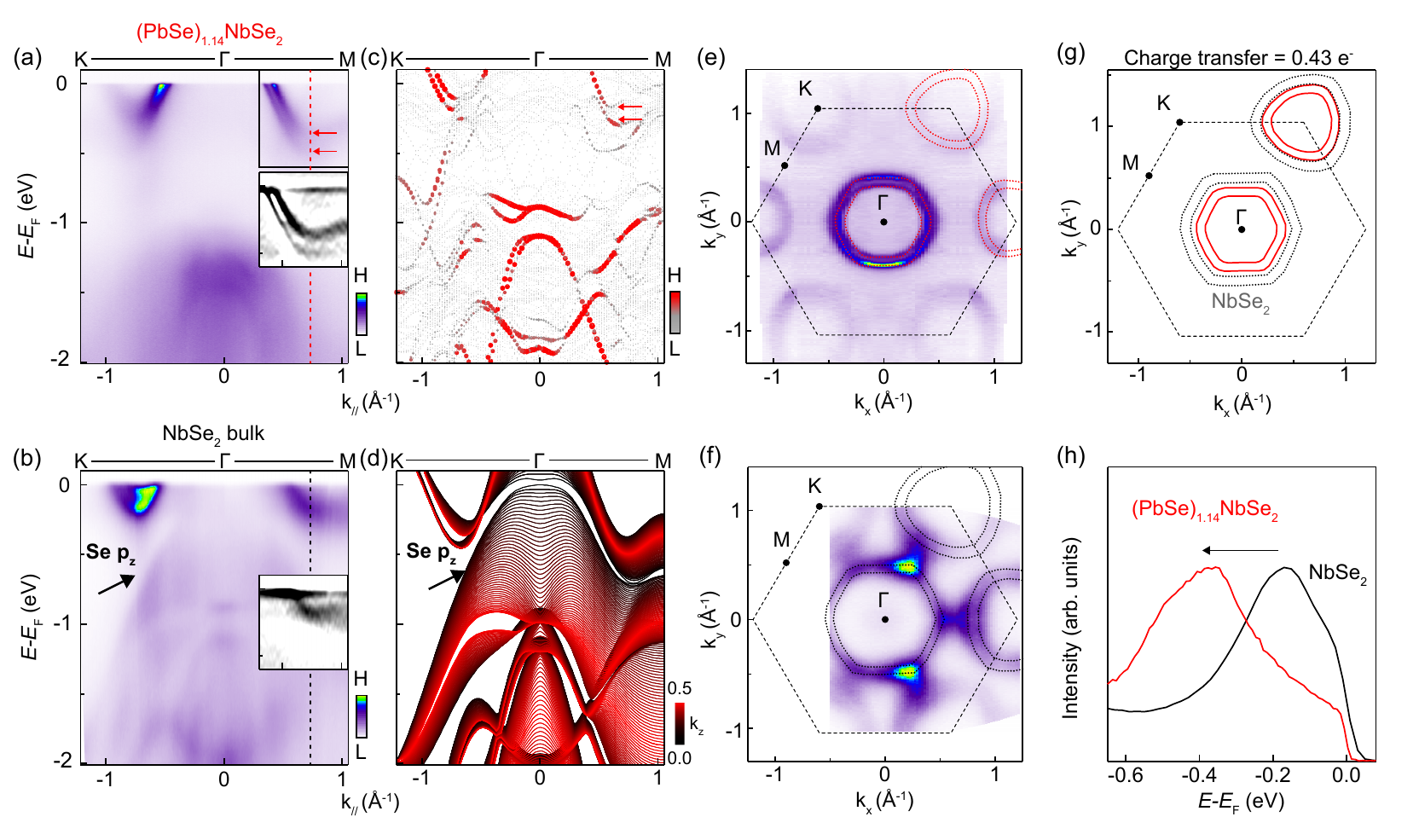}
  \caption{2D electronic structure and charge transfer in (PbSe)$_{1.14}$NbSe$_2$. (a) Dispersion images of MLC (PbSe)$_{1.14}$NbSe$_2$ measured along the K-$\Gamma$-M direction. (b) Dispersion images of bulk NbSe$_2$ crystal measured along the K-$\Gamma$-M direction. The second derivative of the bands near the M point are clearer as shown in the inset.  (c) The unfolded electronic structure of (PbSe)$_{1.14}$NbSe$_2$ projected to the K-$\Gamma$-M direction of NbSe$_2$. The projection weight is represented by the size of the symbols and color. (d) Bulk band structure of NbSe$_2$ projected on the (001) surface Brillouin zone, which shows highly-dispersive Se $p_z$ bands.  (e,f) Symmetrized Fermi surface maps for (PbSe)$_{1.14}$NbSe$_2$ (e) and bulk NbSe$_2$ (f). Red and black dotted curves indicate the hole pockets at the Fermi energy. (g) Extracted Fermi surface pockets for (PbSe)$_{1.14}$NbSe$_2$ (red curves) and bulk NbSe$_2$ (black dotted curves). (h) Extracted energy distribution curves (EDCs) for (PbSe)$_{1.14}$NbSe$_2$ (the red curve) and bulk NbSe$_2$ (the black curve) at momentum positions indicated by red and black dashed lines in (a) and (b).}
  \label{F3}
\end{figure*}

\label{Tex3}

 In order to understand the physics underlying the anisotropic superconductivity in (PbSe)$_{1.14}$NbSe$_2$, ARPES measurements have been performed to reveal the highly 2D electronic structure. Figure 3(a,b) shows a comparison of the experimental electronic structures for (PbSe)$_{1.14}$NbSe$_2$ and bulk NbSe$_2$. 
We note that there are two natural cleaving surfaces, with PbSe and NbSe$_2$ terminations respectively, and they show distinctive core levels (see Fig.~S6, S7 and Table S2 for more details of two terminations in Supplemental Material \cite{supp}). However, the bands near $E_F$ from the Nb $4d$ orbital \cite{IS3,sepz,nb4d}, which  are the main focus, are observed for both terminations and they show overall similar dispersion. A comparison between the misfit compound and bulk NbSe$_2$ shows that the highly-dispersive 3D bands contributed by Se $p_z$ orbital \cite{sepz} in bulk NbSe$_2$ (pointed by black arrows in Fig.~3(b) and Fig.~3(d)) are missing in (PbSe)$_{1.14}$NbSe$_2$ (as shown in Fig.~3(a) and Fig.~3(c)), leading to a large energy separation between the bottom of the conduction band and the top of the valence band. The calculated electronic structure of (PbSe)$_{1.14}$NbSe$_2$ (shown in Fig. 3(c)) agrees qualitatively well with the ARPES results in Fig.~3(a), and they both show distinctive layer-decoupled features that resemble the band structure of monolayer NbSe$_2$ \cite{nbse2yw,lian2018unveiling,mononbse2arpes,dreher2021proximity} (see Fig.~S8, S9  for more details of first-principles calculation in Supplemental Material \cite{supp}). The 2D electronic structure is further supported by the negligible $k_z$ dispersion of
  (PbSe)$_{1.14}$NbSe$_2$ (see Fig.~S10 for more details of $k_z$ dispersion in Supplemental Material \cite{supp}). Therefore, our ARPES measurements together with first-principles calculations show that by expanding the interlayer spacing, the electronic structure of (PbSe)$_{1.14}$NbSe$_2$ becomes layer-decoupled, as indicated by the disappearance of the highly-dispersive Se $p_z$ bands at the $\Gamma$ point at different $k_z$ values. Such 2D electronic structure is consistent with the enhanced superconducting anisotropy in response to the applied magnetic fields. We note while  PbSe bands are absent near $E_F$, similar to previous reports in (PbSe)$_{1.16}$(TiSe$_2$)$_m$ \cite{tise2prl}, the PbSe blocking layer affects the electronic structure from two aspects. Firstly, the PbSe blocking layer plays an important role in breaking the $D_{3h}$ symmetry of the misfit layer compound, which results in the splitting of the Nb $4d$ band along the $\Gamma$-M direction (see Fig.~S11 for more details of D$_{3h}$ symmetry breaking in Supplemental Material \cite{supp}). This is also in contrast to the broken inversion symmetry in monolayer NbSe$_2$, where the Nb $4d$ bands split along $\Gamma$-K direction, while they are still degenerate along $\Gamma$-M direction \cite{mononbse2arpes}. Secondly, it contributes charges to the neighboring NbSe$_2$ as discussed in details below.

 ARPES measurements not only reveal the layer-decoupled electronic structure, but also allow to reveal the charge transfer between PbSe and NbSe$_2$. The carrier concentration can be extracted from the Fermi pockets using the Luttinger theorem \cite{Luttinger_1960} and the charge transfer can be directly extracted by the change of the carrier concentration, as has been recently demonstrated in a different misfit layer compound (PbSe)$_{1.16}$(TiSe$_2$)$_m$ \cite{tise2prl}. The hole pockets of (PbSe)$_{1.14}$NbSe$_2$ shown in Fig.~3(e) are clearly smaller than those of bulk NbSe$_2$ (Fig.~3(f)). This suggests a lower hole concentration, indicating electron transfer from the blocking PbSe layers to active NbSe$_2$ layers. Figure 3(g) shows a comparison of the Fermi pockets for bulk NbSe$_2$ (black dotted curves) and (PbSe)$_{1.14}$NbSe$_2$ (red curves). From the change in the size of the Fermi pockets, a charge transfer of $0.43 ~\pm ~0.02$ electrons per NbSe$_2$ unit cell from the PbSe layer is revealed  (see Fig.~S12 for more details in Supplemental Material \cite{supp}), which is comparable to the value in (PbSe)$_{1.16}$(TiSe$_2$)$_m$ \cite{tise2prl} and (LaSe)$_{1.14}$(NbSe$_2$)$_2$ \cite{leriche2021misfit}. Such charge transfer also results in a band shift of the Nb 4$d$ bands by 250 $\pm$ 20 meV, as shown by the EDCs in Fig.~3(h). These results demonstrate that the PbSe blocking layers not only provide effective protections for the NbSe$_2$ layers and make NbSe$_2$ layers decoupled, but also inject carriers to NbSe$_2$ layer, thereby tuning the carrier concentration of (PbSe)$_{1.14}$NbSe$_2$.
	
	\begin{figure*}[htbp]
  \includegraphics[]{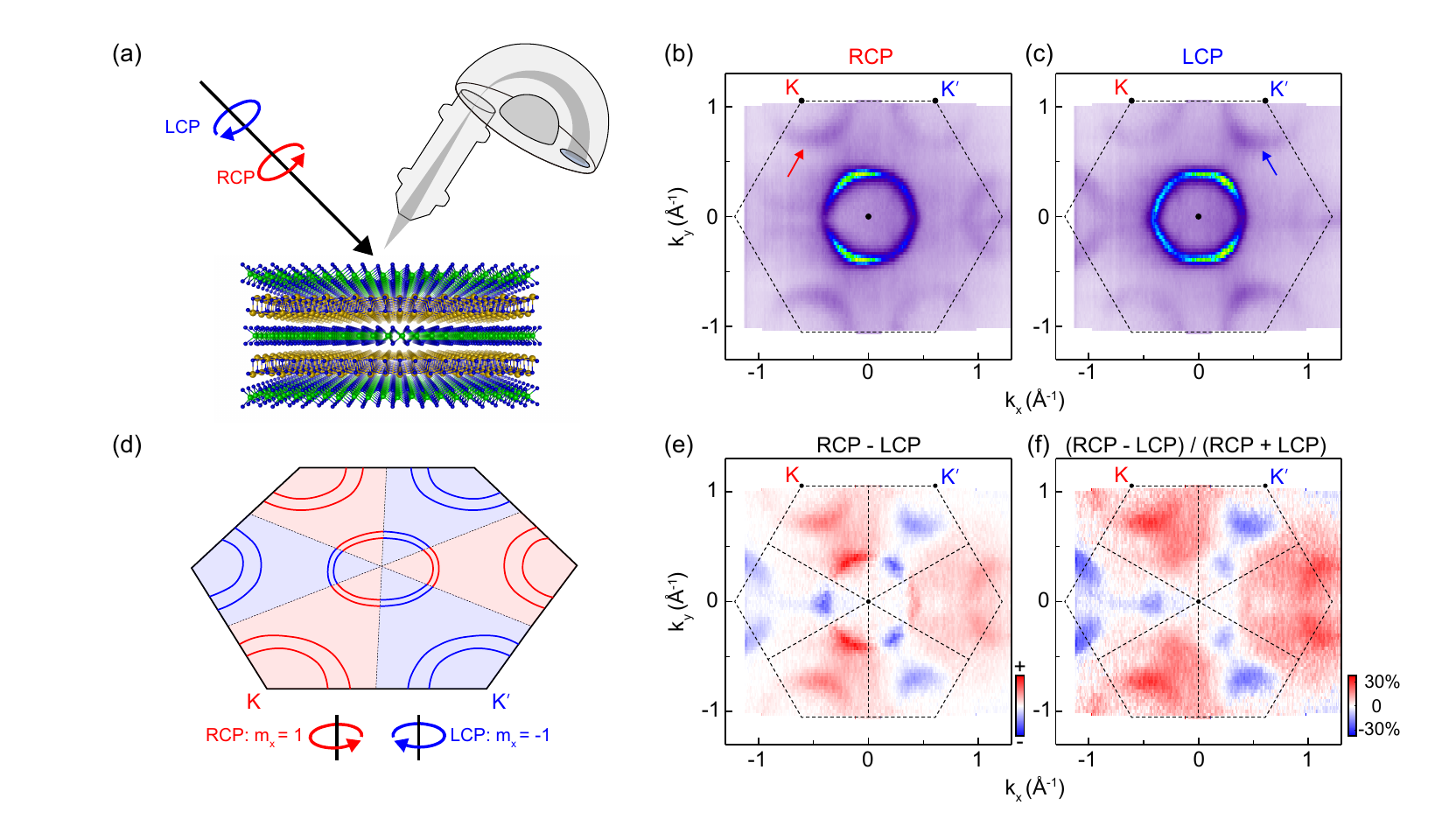}
  \caption{Circular dichroism at different K and K$^{\prime}$ valleys revealed by polarization-dependent ARPES. (a) Schematic drawing of polarization-dependent ARPES measurements with left-handed circular polarization (LCP) and right-handed circular polarization (RCP) light. (b,c) Fermi surface maps measured under RCP (b) and LCP (c) light, respectively. The images are symmetrized with respect to $k_y$ = 0. (d) Schematic drawing of the opposite circular dichroism at K and K$^{\prime}$ valleys. (e) Differential intensity map by subtracting (c) from (b). (f) Circular dichroism map normalized as $(I_{RCP}-I_{LCP}) / (I_{RCP}+I_{LCP})$.}
  \label{F4}
\end{figure*}

\label{Tex4}
 The measured electronic structure confirms that the (PbSe)$_{1.14}$NbSe$_2$ exhibits 2D electronic properties similar to the monolayer NbSe$_2$, yet with a much larger electron doping due to charge transfer from the PbSe blocking layers.  The decoupling of NbSe$_2$ layers suggests that the K and K$^\prime$ valleys are inequivalent due to the inversion symmetry breaking \cite{Xi2015Ising, nbse2spinvalley} (see Fig. S13 for more details of inversion symmetry breaking in Supplemental Material \cite{supp}), which could result in circular dichroism (CD) \cite{mos2pl} and orbital angular momentum (OAM) \cite{Jiang2015Signature}. To search for the OAM and the possible valley pseudospin in (PbSe)$_{1.14}$NbSe$_2$, we perform polarization dependent ARPES measurements as schematically illustrated in Fig.~4(a) (see Fig.~S14 for geometry of CD measurement in Supplemental Material \cite{supp}), where the magnetic quantum numbers for right-handed circular polarization (RCP) and left-handed circular polarization (LCP) lights are opposite \cite{cd_ti_2} as shown in Fig.~4(d). Figure 4(b,c) shows the Fermi surface maps measured by RCP and LCP lights, with clear intensity contrast for the K and K$^\prime$ valleys.  The RCP light enhances the intensity of the K valley as indicated by the red arrow in Fig.~4(b), while the LCP light enhances the K$^\prime$ valley as indicated by the blue arrow in Fig.~4(c). Therefore, the circularly polarized light can selectively couple to these two valleys as schematically summarized in Fig.~4(d), similar to the valleytronics in monolayer MoS$_2$ \cite{mos2pl,mos2pl2} (see Table S3 for comparison of CD data to bulk and monolayer in Supplemental Material \cite{supp}). Such opposite circular dichroism is more clearly revealed in the differential intensity map shown in Fig.~4(e), which is obtained by subtracting Fig.~4(c) from Fig.~4(b). The circular dichroism is further normalized as $(I_{RCP}-I_{LCP}) / (I_{RCP}+I_{LCP})$ \cite{mos2pl} and shown in Fig.~4(f) with an extracted maximum circular dichroism value of 30$\%$. The observation of opposite circular dichroism suggests that there is a significant OAM, since the orbital part of the wave function directly couples to the electric field of the incident light \cite{Jiang2015Signature,cd_ti_2}.  Considering that there is decoupled interlayer coupling in the MLC, the valley contrasted OAM and the strong spin-orbital coupling (SOC) in NbSe$_2$ implies that there are likely different spin textures in different valleys, similar to the spin texture in monolayer NbSe$_2$ \cite{Xi2015Ising}. We note that the correspondence of circular dichroism and spin texture has been revealed on topological insulators \cite{cd_ti_1,cd_ti_5} and Rashba systems \cite{cd_ra_1,cd_ra_4,cd_ra_5}, which also exhibit strong SOC. Here, our results provide the observation of valley-dependent OAM in a MLC system (PbSe)$_{1.14}$NbSe$_2$, which further demonstrates the tailored 2D behavior in the bulk van der Waals superlattice with potential valleytronics applications.

\label{summary}
 In summary, we reveal the 2D electronic structure and anisotropic superconductivity in a natural van der Waals superlattice (PbSe)$_{1.14}$NbSe$_2$. Anisotropic superconductivity with a  transition temperature of  $T_c$ = 5.6 $\pm$ 0.1 K is observed, and opposite circular dichroism at K and K$^\prime$ valleys shows that circularly polarized light can selectively couple with K and K$^\prime$ valleys, thereby suggesting (PbSe)$_{1.14}$NbSe$_2$ as a promising candidate for valleytronics with strong SOC and OAM.  The enhanced superconducting anisotropy compared to bulk NbSe$_2$ is attributed to the reduced dimensionality, which is supported by the increasing interlayer spacing supported by XRD and Raman measurements and the 2D electronic structure revealed by ARPES measurements.
 
 Finally, we would like to discuss the physics implied from the misfit layer compound. The enhanced $H_{c2}$  (more Ising-like superconductivity) is related to the breaking of the inversion symmetry in the misfit compound, similar to the case of monolayer NbSe$_2$ \cite{Xi2015Ising}. The PbSe blocking layer not only induces charges to the neighboring NbSe$_2$ with a charge transfer of 0.43 electrons, but also it plays a role of protecting the NbSe$_2$. Such protection is a possible reason for why the $T_c$  is  higher than monolayer NbSe$_2$, similar to the case of ionic liquid cation intercalated NbSe$_2$ \cite{2022zsynbse2}.  Moreover, the incommensurate PbSe blocking layer also leads to breaking of the D$_{3h}$ symmetry, resulting in a splitting of the bands along the $\Gamma$-M direction, which is distinguished from monolayer NbSe$_2$. 
Our work directly reveals the 2D electronic structure underlying the anisotropy superconductivity in MLC, and provides another pathway to tailor the dimensionality and properties of NbSe$_2$ and other layered materials in addition to ionic liquid cation intercalation \cite{2022zsynbse2}, with properties that are distinctive from the bulk samples and monolayer samples.

\begin{acknowledgments}
\section*{ACKNOWLEDGMENTS}
This work is supported by the National Key R$\&$D Program of China (Grant No.~2020YFA0308800, 2021YFA1400100)  the National Natural Science Foundation of China (Grant No.~12234011, 92250305, 11725418, 12025405, 11874035, 51788104, 51991343 and 21975140),  the National Key R$\&$D Program of China (Grant No.~2018YFA0307100 and 2018YFA0305603), and Fundamental Research Funds for the Central Universities (Buctrc202212). H. Z. acknowledges support from the Shuimu Tsinghua Scholar project and the Project funded by China Postdoctoral Science Foundation (Grant No. 2022M721887). This research used resources of the Advanced Light Source, which is a DOE Office of Science User Facility under contract No. DE-AC02-05CH11231; and the Beamline BL09U of the Shanghai Synchrotron Radiation Facility (SSRF).
\end{acknowledgments}

\subsection{Conflict of interest}
The authors have no conflicts to disclose.


\end{document}